\documentclass[lettersize,journal]{IEEEtran}
\usepackage{amsmath,amssymb,mathtools,mathrsfs,cite}
\usepackage{algorithm,algpseudocode}
\usepackage{array}
\usepackage[caption=false,font=normalsize,labelfont=sf,textfont=sf]{subfig}
\usepackage{textcomp}
\usepackage{stfloats}
\usepackage{url}
\usepackage{verbatim}
\usepackage{graphicx}
\usepackage{color}
\def\BibTeX{{\rm B\kern-.05em{\sc i\kern-.025em b}\kern-.08em
    T\kern-.1667em\lower.7ex\hbox{E}\kern-.125emX}}
\usepackage{balance}

\def \endprf{\hfill {\vrule height6pt width6pt depth0pt}\medskip}

\newcommand{\R}{\mathbb{R}}
\newcommand{\C}{\mathbb{C}}

\newcommand{\e}{e}
\renewcommand{\j}{j}

\newcommand{\vct}[1]{\boldsymbol{#1}}

\newcommand{\<}{\langle}
\renewcommand{\>}{\rangle}

\newcommand{\set}[1]{\mathcal{#1}}

\newcommand{\va}{\vct{a}}
\newcommand{\vb}{\vct{b}}

\newcommand{\vd}{\vct{d}}

\newcommand{\vv}{\vct{v}}
\newcommand{\vw}{\vct{w}}
\newcommand{\vx}{\vct{x}}

\newcommand{\vpsi}{\vct{\psi}}

\newcommand{\setS}{\set{S}}

\graphicspath{{./Figs/}}

\title{Real-time Digital RF Emulation -- I: The Direct Path Computational Model}
\author{C. DeLude, J. Driscoll, M. Mukherjee, N. Rahman, U. Kamal, X. Mao, S. Khan, H. Sivaraman,\\ E. Huang, J. McHarg, M. Swaminathan, S. Pande, S. Mukhopadhyay, and J. Romberg
\thanks{
C. DeLude (email: cdelude3gatech.edu), J. Driscoll, M. Mukherjee, N. Rahman, U. Kamal, X. Mao, S. Khan, S. Pande, S. Mukhopadhyay, and J. Romberg are with the School of Electrical and Computer Engineering at the Georgia Institute of Technology, Atlanta, GA, USA.\newline\indent
M. Swaminathan is currently the department head of Electrical Engineering at The Pennsylvania State University, State College, PA, USA. E. Huang is currently a simulation and control systems engineer with Apple Inc., Cupertino, CA, USA. H. Sivaraman is currently a system software engineer with Nvidia Corp., Santa Clara, CA, USA. They were with the School of Electrical and Computer Engineering at the Georgia Institute of Technology, Atlanta, GA, USA when this project was originally conceived. \newline\indent
J. McHarg is with the Advanced Concepts and Technologies Group at the MIT Lincoln Lab, Lexington, MA, USA. He and his group were crucial to the conception of the direct path model, and provided feedback and test cases throughout all stages of development.\newline\indent
This work was supported by the DARPA Digitial RF Battlespace Emulation (DRBE) program and NIWC Pacific
(N66001-20-C-4001). Any opinions, findings and conclusions are those of the author(s) and do not necessarily reflect the views of DARPA or NIWC Pacific.}
}

\begin{document}

\maketitle

\begin{abstract}
    In this paper we consider the problem of developing a computational model for emulating an RF channel. The motivation for this is that an accurate and scalable emulator has the potential to minimize the need for field testing, which is expensive, slow, and difficult to replicate. Traditionally, emulators are built using a tapped delay line model where long filters modeling the physical interactions of objects are implemented directly. For an emulation scenario consisting of $M$ objects all interacting with one another, the tapped delay line model's computational requirements scale as $O(M^3)$ per sample: there are $O(M^2)$ channels, each with $O(M)$ complexity. In this paper, we develop a new ``direct path" model that, while remaining physically faithful, allows us to carefully factor the emulator operations, resulting in an $O(M^2)$ per sample scaling of the computational requirements. The impact of this is drastic, a $200$ object scenario sees about a $100\times$ reduction in the number of per sample computations. Furthermore, the direct path model
    gives us a natural way to distribute the computations for an emulation: each object is mapped to a computational node, and these nodes are networked in a fully connected communication graph. 
    
    Alongside a discussion of the model and the physical phenomena it emulates, we show how to efficiently parameterize antenna responses and scattering profiles within this direct path framework. To verify the model and demonstrate its viability in hardware, we provide several numerical experiments produced using a cycle level C++ simulator of a hardware implementation of the model. 
\end{abstract}

\section{Introduction}

Modern signal processing has ushered in a new era of RF spectrum usage. From industry to personal use RF processing has become increasingly necessary for everyday life to operate normally. Unsurprisingly, there is an ever-present demand for new systems with higher levels of functionality for use in emerging applications (e.g.\ drones, 5G, automotive radar). Testing and verification of new and existing systems prior to deployment is a necessary but difficult task. This is further exacerbated by new systems often wanting to leverage modern ``machine learning" based processing techniques that require large data sets for training.

As it currently stands, field testing remains the preferable mode of system verification. It is also expensive, slow, and often hard to replicate \cite{matai2012de,bonati2021co}. Each system must be tested in each use case scenario, and with the number of systems and uses increasing this equates to an immense amount of testing. If the system requires training data the fact a field test may be difficult to repeat may be prohibitive. One can imagine that in an airborne radar it can be difficult (if not infeasible) to have pilots run the \emph{exact} same trajectory paths under the \emph{exact} same atmospheric conditions. 

With the clear difficulty and shortcomings of field testing, a much more attractive option for RF testing are so-called ``hardware in-the-loop" emulators. In this paradigm a RF system is plugged directly into emulator, sending into and receiving signals from the device. 
The emulator takes the input signals, processes them in accordance with a specified scenario (e.g.\ flight paths and scattering characteristics of objects, antenna characteristics, clutter models, etc) and returns the result
to the RF system. 
If the emulator can work in real time, it eliminates (or at least greatly reduces) the need to field testing to ``train'' intelligent adaptive radars.
Furthermore, it is straightforward to test many different scenarios, multiple times if needed. 

Although conceptually straightforward, as the outputs of the emulator will just be different combinations of the input signals convolved\footnote{First-order models for Doppler are also straightforward to incorporate into the model.} with different channel responses computed from the scenario, challenges arise in making the emulator accurate, allowing for flexibility in scenarios, and in minimizing latency. 
These problems become more severe as the number of objects involved in the emulation increases, given the super-linear growth of compute requirements with the number of objects. Hence scalability is also of great importance since interesting use cases may involve hundreds of interconnected systems. 
Considering this, organizations such as the Defense Advanced Research Projects Agency (DARPA) and the National Science Foundation (NSF) have launched ambitious programs to usher in a new age of high quality RF emulators \cite{bonati2021co,barcklow2019radio}. To date, all proposed emulator designs use a tapped delay line model where the channel between between interacting objects is modeled as a series of long filter banks\cite{matai2012de,bonati2021co,barcklow2019radio,wickert2001im,borries2009fp,buscemi2011de,jamin2022ac}. As will be discussed in the next section this model is straightforward conceptually and algorithmically, but ultimately scales poorly.

In this paper, which is part I of a two part series, we explore a new computational model based around ``direct path" modeling. In this paradigm each system under test is modeled as a node in fully connected communication graph. The node takes the system under test's signal along with all signals received along the graph edges. It then modifies them in accordance with the channel characteristics of the direct path of propagation between two nodes for a given scenario. Due to the node-centric approach the computational model is easily distributed. As will be discussed below it will give a slightly different result than the TDL model, but is no less accurate. Furthermore, leveraging a mild assumption on the system's RCS characteristics the computations can be factored such that the computational requirements of the model is drastically smaller than that of the TDL model. As a more explicit example of the types of savings we can expect, for a $200$ object scenario the per-sample computational requirement for the direct path model is about $100\times$ less than its TDL equivalent.
 These features, coupled with the flexibility of the model, make it an interesting option to consider in future research and development of RF emulators.

Part II of this series is focused on the practical implementation of this model using application specific integrated circuits (ASICs) or field programmable gate arrays (FPGAs). The ASIC design demonstrates the ability to implement the model in high-bandwidth scenarios, and is largely a more thorough treatment of the design we initially introduced in \cite{mukherjee2023ah}.  The complementary FPGA implementation demonstrates the scalability of the model to scenarios with a larger number of objects. We note that the experiments in part I and a large portion of the experiments in part II use the same base cycle-level C++ simulator. However, the experiments in this paper are meant to validate the models accuracy, whereas the experiments in part II are mainly meant to test the hardware's operational capabilities. The complimentary results in these two papers, and in particular the existence of the purpose built hardware necessary for implementation, demonstrate that the model is viable in a practical sense.

\section{RF channel modeling}
\begin{figure*}[t]
    \centering
    \begin{tabular}{cc}
         \includegraphics[width = .65\textwidth]{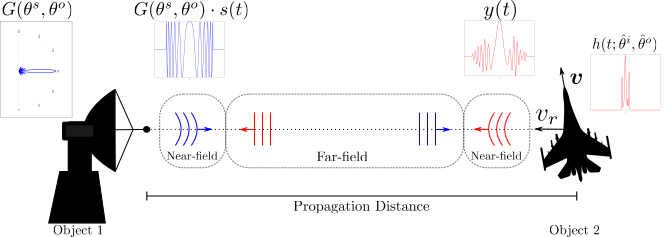} &  \includegraphics[width = .2\textwidth]{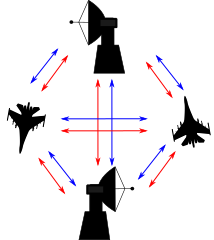}\\
         (a) Two-object scenario & (b) Multi-object scenario 
    \end{tabular}
    \caption{\small \sl (a) For a two object scenario where one of the objects is a pure reflector we must take into account several physical phenomena. Highlighted here is the angular-dependent antenna gain, angular/object-dependent scattering response (or RCS), scenario-dependent relative motion, and scenario-dependent propogation distance. (b) When we introduce multiple objects to the scenario we must model all of these interactions simultaneously to form an accurate channel model.}
    \label{fig:rfchannel}
\end{figure*}

In order to build an accurate RF channel emulator we must first establish an RF channel model that we wish to emulate. Ideally this model would incorporate every possible physical phenomena that an electromagnetic wave may encounter while propagating through some media, but this is hugely impractical. Therefore it is standard to model only the physical phenomena considered to be significant such as propagation loss, antenna gains, scattering profiles, etc. \cite{skolnik2008ra,yin2016pro}. This discussion is meant to emphasise that in terms of model granularity there is almost no limit to the complexity with which we can model an RF channel. In this section we will overview what physical phenomena a standard emulator captures. We begin by describing a simplified two-object example, using the case scenario shown in  Figure~\ref{fig:rfchannel}(a) as a working example. We then show how the model extends to an arbitrary number of objects. Additional discussion is given to scattering profile approximations, which are critical to both the TDL model and our proposed method. Finally, we discuss the computational complexity of the TDL model implementation.

\subsection{Simplified two-object model}
\textbf{Antenna Gain}. Our model starts with a signal being radiated or received by an antenna at a spherical angle $\theta = (\varphi,\vartheta)$ where $\varphi$ denotes azimuth and $\vartheta$ denotes elevation. At the same time the radiator is being steered (either physically or electronically) at a similarly defined angle $\theta^s$. We denote the radiators angular-dependent gain by the function $G:\setS^2\times \setS^2  \mapsto \mathbb{C}$ that multiplies the signal upon receive or prior to transmission. Technically this function depends on frequency, but we assume that the variation across the bandwidth of signals under test is negligible. This function captures the fact that the signal may not, and likely is not, radiated by an isotropic antenna. Due to reciprocity \cite{stumpf2018el} whether $\theta$ is the incoming or outgoing angle does not change the response at that angle. Nevertheless, for completeness we will use $\theta^i$ and $\theta^o$ to denote the incoming and outgoing angle respectively. For our two object example this means that the signal that leaves object 1 in the direction of object 2 is $G(\theta^s,\theta^o)\cdot s(t)$.

\textbf{Path loss}. The signal then propagates from the ``near-field" to the ``far-field," where the far-field is the propagation regime where the path loss follows an inverse square law (among several other simplificationss). This regime is generally accepted to occur when the propagation distance $d$ far exceeds the signals maximum wavelength and the size of the radiator \cite{rappaport2003mo}. This allows us to model the path loss as a function $C:\mathbb{R}\mapsto\mathbb{R}$ dependent only on the propagation distance and is $\propto \frac{1}{d^2}$\footnote{This too also technically depends on frequency, but again the fluctuation is small over the bandwidth of $s(t)$.}. 

\textbf{Plane wave}. A fundamental assumption is that the interacting objects are far enough away to admit a plane wave approximation e.g.\ where the spherical sector of the radiating wave can be modeled as a plane impinging on the object. This is a completely standard approximation, and is highly accurate for most reasonably spaced objects \cite{bohagen2009on}. This allows us to model the reflecting objects scattering response as a filter parameterized by the angles of incidence as opposed to some more complicated space-time operator. 

\textbf{Relative velocity}. A critical phenomena we must model is relative velocity between objects, which will incorporate several approximations. A thorough summary of these approximations and why they are generally good is given in Appendix~\ref{sec:apdx_doppler}, and here we outline the main points. The first of these being that the radial component of the velocity is the only part that impacts the signal. To formalize this, consider arbitrary objects $n$ and $n'$ at positions $\{\vx_n,\vx_{n'}\}$ traveling at instantaneous velocities $\{\vv_n, \vv_{n'}\}$ respectively. The displacement vector from object $n$ to $n'$ is given by $\vd_{n,n'} = \vx_{n'}-\vx_n$, where we use the convention $\vd_{n,n'}$ to mean the displacement from the $n$th object to the $n'$th object. Similarly the relative velocity is defined similarly with $\vv_{n,n'} = \vv_{n'}-\vv_n$. The relative velocity component aligned along the radial displacement vector is then
\begin{align*}
    v_r = \frac{\< \vd_{n,n'}, \vv_{n,n'} \>}{\|\vd_{n,n'}\|_2}.
\end{align*}
We then make the approximation that (neglecting $G(\theta^s,\theta^o)$ and $C(d)$ for a moment) for a signal $s(t)$  sent from $n$ to $n'$ the velocity dilates the signal to $s\left (t-t\frac{v_r}{c}\right)$ where $c$ is the speed of light. This no longer an LTI system, but we can remedy this inconvenient fact with another approximation. If $s(t)$ is a $\Omega$-bandlimited signal modulated to a carrier frequency $f_c$ such that its spectrum is supported on the inverval $[f_c-\Omega,f_c+\Omega]$, then 
\begin{align*}
    s\left (t-t\frac{v_r}{c}\right) \approx e^{-j 2 \pi f_c \frac{v_r}{c} t}s(t).
\end{align*}
So, instead of dilating our signal, we can approximate the effects of relative motion via a modulation. The modulation frequency due to the reative motion $-f_c \frac{v_r}{c}=-\frac{v_r}{\lambda}$ is often called the Doppler frequency \cite{ballot1845ak}.  Going back to our two-object example, the signal received by the object 2 is
{\small\begin{align*}
    s_{1,2}(t) = G(\theta^s,\theta^o)\cdot C\left(\|\vd_{1,2}\|_2\right) \cdot e^{-j 2 \pi f_c \frac{\< \vd_{1,2}, \vv_{1,2} \>}{c\|\vd_{1,2}\|_2} t} \cdot s(t-\tau_{1,2}),
\end{align*}}
where $\tau_{1,2}$ denotes the propagation delay $\tau_{1,2} = \|\vd_{1,2}\|_2/c$. Moving forward we will use $\tau_{n,n'}$ to denote the propagation delay from object $n$ to object $n'$

\textbf{Scattering profile}. When the a signal impinges on an object it is subject to the object's scattering profile e.g.\ the targets response to a wave incident at a given angle. When the signal is a plane wave the interaction can be modeled as an angular-dependent filter. The filter $h(t;\theta^i,\theta^o)$ admits two angle parameters, an input angle $\theta^i$ and output angle $\theta^o$. This inclusion of differing input and output angles is necessary to account for bistatic situations where the transmitter and receiver are not co-located. In our model we define this scattering respone with respect to the phase center of the object. Hence, in the above example the signal received at the phase center of object 2 is $h(t;\hat\theta^i,\hat\theta^o) \star s_{1,2}(t)$ \cite{potter1995ag,chance2022di}. Note that in this simplified example $\hat\theta^i = \theta^o$ and $\hat\theta^o$ is the antipodal angle of $\hat\theta^i$ since the receiver and transmitter are co-located.

\textbf{Absorption and emission}. Though perhaps a subtle point our model assumes a received signal is absorbed and then re-emitted, which is the proper way to model electromagnetic wave interactions \cite{ditchburn2011li}. Upon re-emission the signal incurs a complementary Doppler modulation, path loss, propogation delay, and antenna gain prior to reaching an adjacent object. With this in mind we can write the signal received by object 1 as 
{\small\begin{align*}
    y(t) =G(\theta^s,\theta^i) \cdot C\left(\|\vd_{2,1}\|_2\right)&\cdot e^{-j 2 \pi f_c \frac{\< \vd_{2,1}, \vv_{2,1} \>}{c\|\vd_{2,1}\|_2} (t-\tau_{2,1})}\\
    &\cdot h(t;\hat\theta^i,\hat\theta^o) \star s_{1,2}(t-\tau_{2,1}),
\end{align*}}
where $\theta^i = \hat\theta^o$. This is simply the signal recevied at the phase center of object 2 re-emitted in the direction of object 1. The additional $G(\theta^s,\theta^i)$ is due to the assumption that the signal goes to the receiver, but in principle object 1 also has a scattering response. Thererfore in addition to the recieved signal and the transmitted signal, it could also reflect a portion of the re-emitted signal back to object 2. This gives rise to the fast decaying, but observable, ``ringing" between objects. 

\subsection{Modeling multiple objects}

Now that we have established the physical phenomena we wish to model, and shown how it fits into a scenario, we can easily extend it to the more interesting case of multiple objects interacting. This case, depicted in Figure~\ref{fig:rfchannel}(b), incorporates several systems; all sending and receiving signals, either actively through radiators or passively through reflections. For our discussion we will assume that there are $N$ interacting objects, and for conciseness we will focus on the $m$th object where $m\in \{1,2,\dots,N\}$.

In the scenario, the $m$th object has its own scattering profile $h_m(t;\theta^i,\theta^o)$, radiator $G_m$, position $\vx_m$, and velocity $\vv_m$. An object sees $N$ different signals labeled as $s_1(t), s_2(t),\dots,s_N(t)$, where $s_m(t)$ is the signal generated by the transmitter and the remaining $N-1$ signals are received from the adjacent objects. The $N-1$ input received signals are combined along with the transmit signal into $N-1$ different outputs. The input signals are assumed to arrive at angles $\{\theta_n^i\}_{n\neq m}$, and the outputs are sent back to their respective objects at angles $\{\theta_n^o\}_{n\neq m}$. Of course, for a match pair $\theta_n^i$ is the antipodal angle of $\theta_n^o$. With these definitions in place the output signal from object $m$ to object $\ell$ is given by
{\small\begin{align}
    \label{eq:tdl_transmit}
    \nonumber
    y_{m,\ell}'(t) &= G_m(\theta_m^s,\theta_\ell^o) s_m(t-\tau_{m,\ell})\\
    \nonumber
    & \qquad\qquad\qquad\qquad+ \sum_{\substack{n=1\\ n \neq m}}^N h_m(t;\theta_n^i,\theta_\ell^o)\star s_n(t-\tau_{m,\ell})\\
    y_{m,\ell}(t) &= C(\|\vd_{m,\ell}\|)\cdot e^{-j 2 \pi f_c \frac{\< \vd_{m,\ell}, \vv_{m,\ell} \>}{c\|\vd_{m,\ell}\|_2} (t-\tau_{m,\ell})}  \cdot y_{m,\ell}'(t).
\end{align}}
Of course, if the object is passive then the transmit term above can be neglected. Similarly, if the scattering profile of the object is negligible then we can set $h_m(t;\theta^i,\theta^o)=0$. Whether the object is a transmitter, receiver, both, or neither  \eqref{eq:tdl_transmit} fully describes the one-way channel from object $m$ to object $\ell$ under our model. If object $m$ is equipped with a receiver and we want to read a signal off of the reciever, then the $m$th receivers signal is given by 
\begin{align}
    \label{eq:tdl_rec}
    r_m(t) = \sum_{\substack{n=1\\ n \neq m}}^N G_m(\theta^s_m,\theta_n^i)\cdot s_n(t-\tau_{n,r}).
\end{align}
The $\tau_{n,r}$ are temporal offsets, implicitly dependent on $\theta_n^i$, that compensate the receive signal to account for the fact the receiver is likely not at the object's phase center. They can be calculated analogously to the point scattering temporal offsets discussed in the next section. So whereas \eqref{eq:tdl_transmit} is modeling the channel along a given path \eqref{eq:tdl_rec} allows us to sample the signal at a particular receiver.

\subsection{Scattering profile approximations}
\label{sec:scatteringprofile}
\begin{figure}[h]
    \centering
    \begin{tabular}{cc}
         \includegraphics[width = .25\textwidth]{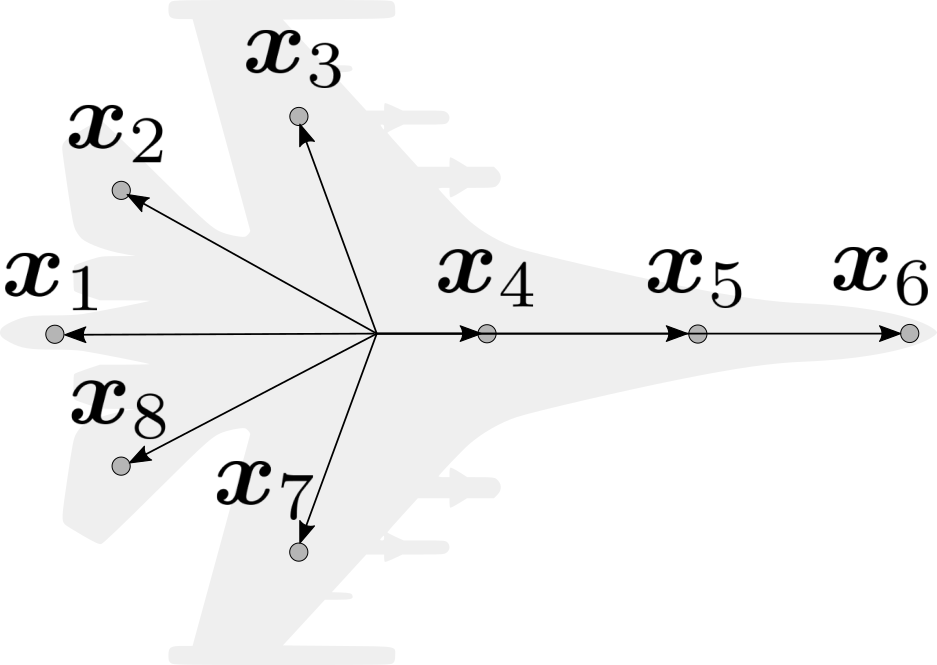} \\
         (a)\\
         \includegraphics[width = .45\textwidth]{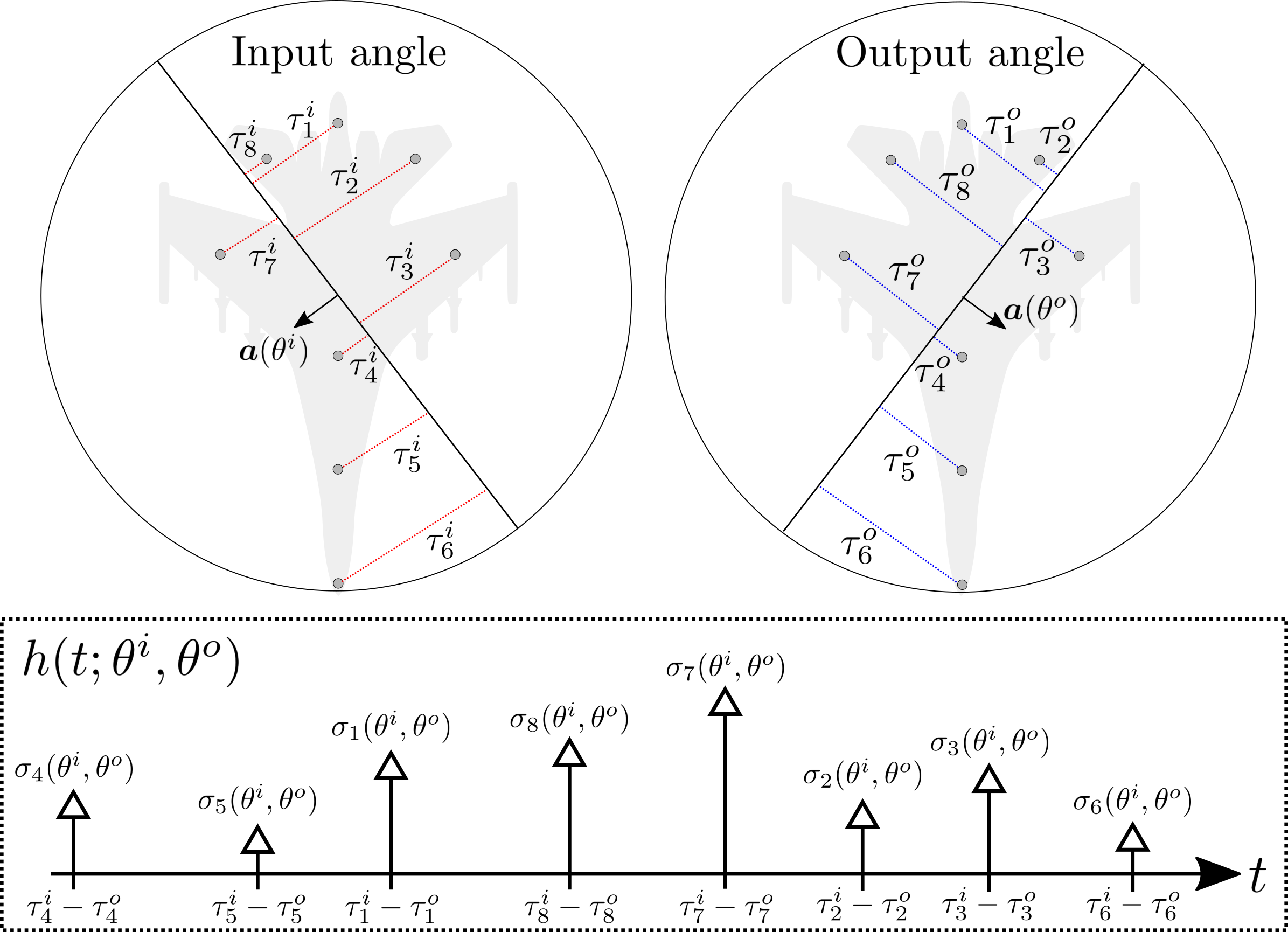}  \\
         (b) 
    \end{tabular}
    \caption{\small \sl (a) Each scatterer has location $\vx_k$ in the local coordinate system.  (This is depicted here in 2D to simplify the sketches.)  (b) An incoming plane wave at angle $\theta^i$ induces a set of delays $\tau^i_k$ relative to the virtual center of the object.   The outgoing response plane wave at angle $\theta^o$ induces delays $-\tau^o_k$. These delays coupled with the bistatic complex weightings $\{\sigma_k(\theta^i,\theta^o)\}_{k=1}^K$ are used to generate the scattering response filter $h(t;\theta^i,\theta^o)$.} 
    \label{fig:sepscattering}
\end{figure}
A fundamental assumption that emulators generally leverage is that the scattering profile admits a point scattering approximation. In this paradigm the scattering response of an object can be modeled as a set of discrete scattering points at fixed locations. Though seemingly restrictive, this is accepted to be a sufficient model for most frequency ranges of interest \cite{rihaczek2000th}.

To formalize this, let $\{\vx_k\}_{k=1}^K$ for $\vx_k\in\mathbb{R}^3$ be the locations of a set of scatters used to approximate the response of an object as shown in Figure~\ref{fig:sepscattering}(a). For a signal arriving at $\theta^i$ and being reflected at $\theta^o$ the relative temporal offset of the signal with respect to the phase center of the object is
\begin{align*}
    \tau_k^i = \vx_k^T\va(\theta^i),~\tau_k^o = \vx_k^T\va(\theta^o),~\va(\theta) =
    \frac{1}{c}
    \begin{bmatrix}
        \cos\varphi\sin\vartheta\\
        \sin\varphi\sin\vartheta\\
        \cos\vartheta
    \end{bmatrix},
\end{align*}
as shown in Figure~\ref{fig:sepscattering}(b-c). With these definitions in place the scatterer seperable model in its most general form is
\begin{align}
    \label{eq:pointscattering}
    h(t;\theta^i,\theta^o) = \sum_{k=1}^K \sigma_k(\theta^i,\theta^o)\delta(t-\tau_k^i+\tau_k^o),
\end{align}
where $\sigma_k:\setS^2\times\setS^2 \mapsto \mathbb{C}$ is an angular dependent complex weighting. The intuition is that the \eqref{eq:pointscattering} shifts the signal from the object origin to each of the scatterers (corresponding to the delays $\tau_k$). Then these scatterers re-emit the signal after scaling it by $\sigma_k(\theta^i,\theta^o)$, creating a signal that is equivalent to being emitted from the scattering point.

This model has been widely studied in literature under varying modeling assumptions. For example, in \cite{konovalyuk2014pa,chance2022di} examine using \eqref{eq:pointscattering} under an isotropic point scattering assumption where $\sigma_k(\theta^i,\theta^o)=\sigma_k$. This is a perfectly sufficient assumption for most applications, including imaging, and has the benefit of being relatively straightforward to understand and use. For an anisotropic scattering assumption (e.g.\ when the complex weighting is angle dependent) \cite{ash2015wi} gives a thorough overview of various modeling techniques. A popular method is to model $\sigma_k(\theta^i,\theta^o)$ as the angular response from a known object geometry such as a corner, plate, etc.\ as discussed in \cite{potter1995ag}. We note that these references primarily focus on modeling $\sigma_k(\theta^i,\theta^o)$ in a monostatic setting where the receiver and transmitter are collocated. Though the same point scattering model still holds, the bistatic response where $\theta^i\neq\theta^o$ has been studied to a lesser extent. Nevertheless the approach is similar to the monostatic case with \cite{jackson2008pa} providing expressions for modeling the response of known object geometries which could be used in tandem with the approach in \cite{potter1995ag} to generate a bistatic profile. 

With this point scattering approximation in place the output signal from object $m$ to object $\ell$ given in \eqref{eq:tdl_transmit} becomes
{\small
\begin{align}
    \label{eq:tdl_transmit_nonsep}
    \nonumber
    y_{m,\ell}'(t) &= G_m(\theta_m^s,\theta_\ell^o) s_m(t-\tau_{m,\ell})\\
    \nonumber
    &\qquad + \sum_{\substack{n=1\\ n \neq m}}^N \sum_{k=1}^K \sigma_{m,k}(\theta^i_n,\theta^o_\ell) s_n(t-\tau_{n,k}-\tau_{m,\ell} +\tau_{k,\ell} )\\
    y_{m,\ell}(t) &= C(\|\vd_{m,\ell}\|)\cdot e^{-j 2 \pi f_c \frac{\< \vd_{m,\ell}, \vv_{m,\ell} \>}{c\|\vd_{m,\ell}\|_2} (t-\tau_{m,\ell})}  \cdot y_{m,\ell}'(t).
\end{align}}
Hence the objects scattering profile is now approximated by a $K$ point FIR filter. Of course, the filter changes with each input and output angle pair.

\subsection{TDL model and computational complexity }

As may be apparent from the above discussion, even modeling a simple two object scenario is complicated. We must calculate and keep track of a plethora of variables and parameters while successively applying them to multiple signals. As the number of interacting objects in a scenario increases, this problem is only exacerbated. Hence, computational complexity and how it scales is of keen interest in the development of new models. Here we briefly examine the existing methodology for channel simulation, the TDL model, and show how it scales with the number of objects.

The TDL model can be viewed as a direct implementation of \eqref{eq:tdl_transmit_nonsep} and \eqref{eq:tdl_rec}. Generally the RF signal is digitized prior to being input to the emulator, with all procedures related to signal manipulation being handled in the digital domain \cite{matai2012de,bonati2021co,barcklow2019radio,borries2009fp}. Essentially the summation term in \eqref{eq:tdl_transmit_nonsep} becomes a possibly very long (but reasonably sparse) filter with $O(NK)$ non-zero terms in it.  Note that the digitized signal must be buffered in accordance with the length of the filter, which in turn is determined by the maximum operational range of emulation. Each object produced $N$ different outputs, hence the per-sample complexity for each object is $O(N^2K)$. It follows that the total sample complexity in the TDL model is $O(N^3K)$, and furthermore if the buffer length is $L$ the per object memory complexity is $O(NL)$. If the objects are spaced very far apart in the scenario $L$ can be potentially very large. For example if the round trip distance is $50$ km and the sampling frequency is $f_s=2.5$ GHz, then $L > 4.16\cdot 10^5$. 

Though this is not a particularly favorable computational or memory scaling, it is the standard and accepted way of emulating an RF channel. Efforts to reduce the hardware requirements is largely limited to designing computational clusters \cite{buscemi2011de} or leveraging the sparse structure of the filters \cite{jamin2022ac}. Hence their ultimately is not much we can do to improve upon the baseline requirements of TDL. However, In the next section we introduce a subtly different computational model that is more efficient in both memory and computation.

\section{Direct Path Model}
\label{sec:directpath}

Our computationally efficient approach to RF channel emulation is based around a direct path model. As will be discussed in this paradigm each object in the scenario is modeled as a computational node in a fully connected communication graph. The computational savings arise from leveraging a mild but standard modeling assumption on the objects scattering profile. This allows us to carefully factor computations in a manner that reduces the total sample complexity from $O(N^3K)$ in the TDL model to $O(N^2K)$ computations with similar savings in terms of memory complexity. This section will overview this model, beginning with the scattering profile assumption.

\subsection{Separable scattering profile}

The discussion in Section~\ref{sec:scatteringprofile} is meant to emphasize that with the point scattering model in hand there are several viable ways to model the complex weighting factor. Though the richest model is to calculate $\sigma_k(\theta^i,\theta^o)$ directly for each angle pair, dropping the angle dependence entirely is perfectly adequate for many applications. Our direct path model will explicitly leverage a different approach taken in \cite{huang2021an,huang2022sp}, which showed that the bistatic response of an object can be well approximated under a separable model where $\sigma_k(\theta^i,\theta^o) = \alpha_k(\theta^i)\beta_k(\theta^o)$. Moving forward with our formulation we will assume that the scattering profiles are bistatic and separable. Hence \eqref{eq:pointscattering} now takes the form
\begin{align}
    \label{eq:pointscatteringsep}
    h(t;\theta^i,\theta^o) = \sum_{k=1}^K \alpha_k(\theta^i)\beta_k(\theta^o)\delta(t-\tau_k^i+\tau_k^o).
\end{align}
Although, to reiterate, the richest model uses a non-separable form of $\sigma_k$, we will see that enforcing separability comes with tremendous computational benefits. Hence there is a tradeoff in terms of our modeling ability. However, given the simplicity of most scattering models and the comparative richness of the separable model this is a modest tradeoff.

\subsection{Nodal compute structure}
\begin{figure}[h]
    \centering
    \begin{tabular}{cc}
         \includegraphics[width = .4\textwidth]{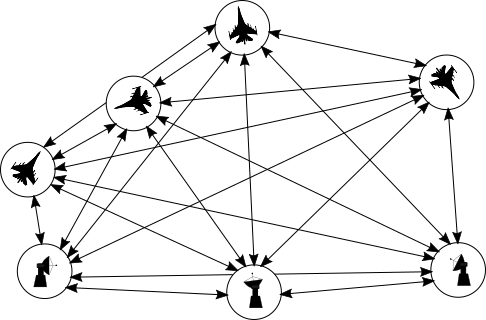} \\
         (a) Nodal representation\\
         \includegraphics[width = .4\textwidth]{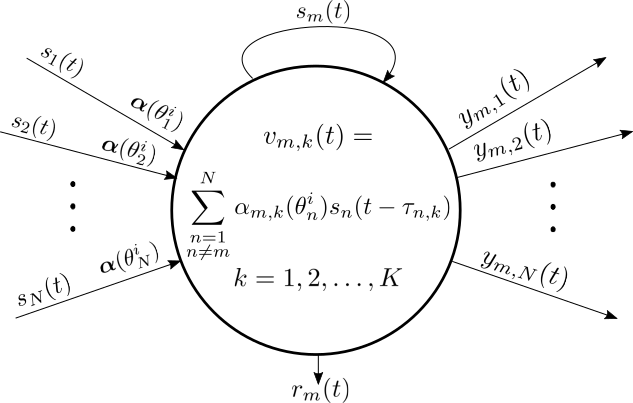}\\
          (b) Computational node 
    \end{tabular}
    \caption{\small \sl (a) The direct path model represents each object in a scene as a node in fully connected communication graph. (b) within each node all $N-1$ signals from adjacent nodes are received and combined into a series of intermediate signals $\{v_k(t)\}_{k=1}^K$. These intermediate signals are then combined in different ways and transmitted transmitted as outputs $\{y_n(t)\}_{n=1}^N$ to adjacent nodes.}
    \label{fig:dpmodel}
\end{figure}
In the direct path model objects in a scenario are modeled as nodes in a communication graph regardless of the kind of object (see Figure~\ref{fig:dpmodel}(a)). So, regardless of if the object is a passive reflector or a base station sending and receiving signals it is treated as a node. The $m$th node receives $N$ input signals $\{s_n(t)\}_{n=1}^N$, reserving $s_m(t)$ for the object's transmitter, and outputs $N-1$ outputs $y_n(t)$ to the adajcent nodes. Each node has an additional output $r_m(t)$ that acts as the receiver output \eqref{eq:tdl_rec} if the object has a receiver.

As previously discussed, we assume (for good reasons) that the scattering response of each object is well approximated by the separable point scatter model described by \eqref{eq:pointscatteringsep}. Under this model the channel response response given in \eqref{eq:tdl_transmit} becomes 
{\small
\begin{align}
    \label{eq:compute_nonfactor}
    \nonumber
    y_{m,\ell}'(t) &= G_m(\theta_m^s,\theta_\ell^o) s_m(t-\tau_{m,\ell})\\
    \nonumber
    &+ \sum_{\substack{n=1\\ n \neq m}}^N \sum_{k=1}^K \alpha_{m,k}(\theta_n^i)\beta_{m,k}(\theta_\ell^o)s_n(t-\tau_{n,k}-\tau_{m,\ell} +\tau_{k,\ell} )\\
    y_{m,\ell}(t) &= C(\|\vd_{m,\ell}\|)\cdot e^{-j 2 \pi f_c \frac{\< \vd_{m,\ell}, \vv_{m,\ell} \>}{c\|\vd_{m,\ell}\|_2} (t-\tau_{m,\ell})}  \cdot y_{m,\ell}'(t).
\end{align}}
We note that in general $\tau_{m,\ell}\gg|\tau_{k,\ell}-\tau_{n,k}|$ due to the separation of objects assumed to being larger than the objects themselves. Hence causality of the system is not violated. The key step in efficiently implementing the direct path model is a factorization of \eqref{eq:compute_nonfactor}. Define $K$ intermediate signals
\begin{align}
    \label{eq:intermediate}
    v_{m,k}(t) = \sum_{\substack{n=1\\n\neq m}}^N \alpha_{m,k}(\theta_n^i)s_n(t-\tau_{n,k})
\end{align}
for $k=1,2,\dots,K$ that are held by the $m$th node. With these intermediate signals in hand we can form an output to the $\ell$th node by taking 
{\small
\begin{align}
    \label{eq:compute_factor}
    \nonumber
    y_{m,\ell}'(t) &= G_m(\theta_m^s,\theta_\ell^o) s_m(t-\tau_{m,\ell})\\
    \nonumber
    &\qquad\qquad+ \sum_{k=1}^K \beta_{m,k}(\theta_\ell^o)v_{m,k}(t-\tau_{m,\ell} +\tau_{k,\ell} )\\
    y_{m,\ell}(t) &= C(\|\vd_{m,\ell}\|)\cdot e^{-j 2 \pi f_c \frac{\< \vd_{m,\ell}, \vv_{m,\ell} \>}{c\|\vd_{m,\ell}\|_2} (t-\tau_{m,\ell})}  \cdot y_{m,\ell}'(t).
\end{align}}
This style of operation is depicted graphically in Figure~\ref{fig:dpmodel}(b) the $N-1$ input signals are collapsed into $K$ intermediate signals held by the node. These intermediate signals are then manipulated to form the $N-1$ outputs. This factorization of operations forms the computational backbone of the direct path model.

\subsection{Computational complexity and distributed processing}

We will now discuss how this factorization greatly improves upon the TDL model in terms of computational complexity. We again assume that the signal is digitized prior to computation with the signals sent to adjacent nodes via some communication network. The $m$th node receives $N-1$ signals that are then formed into $K$ intermediate signals \eqref{eq:intermediate} requiring $O(NK)$ operations. Each intermediate signals is placed in a buffer of length $L$, requiring a memory complexity of $O(KL)$ per node. Forming an output requires combining $K$ intermediate signals with the transmit signal. We must do this $N$ times to form the outputs, requiring an additional $O(NK)$ operations. So a single node has a computational complexity of $O(NK)$ and a memory complexity of $O(KL)$. The total computational complexity of an $N$ object system is then $O(N^2K)$.

To make these advantages more clear, consider an emulator modeling $N=200$ interacting objects. Let each object's scattering profile by modeled by $K=16$ scattering points. We apply the respective delays via a $4$-tap FIR filter such that the response at a fixed angle can be represented by a $64$-tap FIR. If we were to implement \eqref{eq:compute_nonfactor} directly via a TDL model, then producing a single sample output from the channel and receive path requires $\approx 5.07\cdot 10^8$ operations. In contrast, if we use the factorized implementation given by \eqref{eq:compute_factor} then producing a single sample output requires $\approx5.25\cdot 10^6$ operations. Hence we have reduced the computational requirements by nearly a $100\times$ factor.

This alone is a great improvement over the TDL model. However, and additional benefit is that the system is distributed in a natural way. Each computational node essentially runs locally, with minor knowledge of adjacent nodes required. The main requirement to distribute the system is a suitable communication network between channels. A node can be built independently and added to emulation system by integrating into this network. This distribution concept and the factored direct path model are implemented in part II of this paper, in both an ASIC and FPGA version. The ASIC implementation is also discussed in \cite{mukherjee2023ah}, but in less rigorous detail. Ultimately the results in these supplementary works demonstrate that the concepts discussed in this section can be implemented practically.

\section{Technical considerations}
\label{sec:techcon}

The above discussions on modeling require us to manipulate a signal in several ways in order to properly emulate an RF channel. This section will elaborate on how to perform some of these operations. In particular, we will discuss design considerations in applying fractional delays to a sampled signal and how to efficiently parameterize the functions $G$ and $\sigma_k(\theta^i,\theta^o)$. These will ultimately be used in the numerical hardware simulations in Section~\ref{sec:experiments}.

\subsection{Fractional delay filtering}
In most conceivable implementations of the direct path model the signal is sampled close to the RF interface. This signal, sampled at a rate of $f_s$, is then digitally manipulated at each computational node. As is apparent from \eqref{eq:intermediate} and \eqref{eq:compute_factor} several delay operations must be performed on the intermediate and output signal. We could limit the types of delays to integer sample shifts, but this will induce substantial distortion in the scenarios response and dynamics. Therefore, it is necessary to apply interpolation filters to induce a fractional delay.

This style of filtering is completely standard, but still requires careful consideration. We focus on an $R$-tap FIR filter design and examine two coefficient generation methods; quadratic spline and Legendre polynomial interpolation. We define the following two metrics to quantify performance:
\begin{itemize}
    \item[] \underline{Delay Accuracy} -- Maximum group delay deviation (in ns) from delay setting taken over all frequencies and all delay settings.
    \item[] \underline{Amplitude Ripple }-- Difference between maximum and minimum amplitude response taken over all frequencies and delay settings. 
\end{itemize}
The term ``delay setting" simply means the number of fractional delays we discretize the normalized range $(0,1)$ to. Something we can do to improve these metrics is to oversample the signal. This is because the highest amounts of distortion in the group delay and amplitude are at the highest frequencies. Hence, if we oversample, then the occupied portion of the signals band does not interact with the high distortion region of the filter. Since the ideal fractional delay filter is infinite in length it is not surprising that performance can also be improved by increasing the number of taps. So, two ways we can meet some performance threshold is by increasing the sampling frequency or increasing the number of taps $R$.

Two examine this tradeoff more thoroughly Table~\ref{fig:filtperf} displays the delay accuracy and amplitude ripple for the previously mentioned spline and Legendre filter for varying $R$ and oversampling percentages. The signal is assumed to have a bandwidth of $2$ GHz, hence a delay accuracy of $.1$ ns equates to a range accuracy of $\sim3$ cm. It is apparent from these results that oversampling allows for shorter filters to perform as well or better than longer filters that do not have the benefit of being oversampled. Moving forward with our numerical experiments in Section~\ref{sec:experiments} we will use 4-tap spline filters subject to $25\%$ oversampling.
\begin{table}[h]
    \centering
    \caption{\small \sl (a) Amplitude ripple and (b) delay accuracy of spline and Legendre based interpolation filters for a variety of filter lengths and oversampling rates. The oversampling percentage is defined with respect to the signals baseband bandwidth of $2$ GHz.}
    \label{fig:filtperf}
    \begin{tabular}{cc}
    (a) Amplitude ripple\\
    \small{
    \begin{tabular}{||c|c|c|c|c||}
        \hline
         Filter & 20\% & 25\% & 30\% & 33\%  \\
         \hline\hline
         Legendre - 4 Tap & 0.62 & 0.55 & 0.49 & 0.45\\
         \hline
         Legendre - 8 Tap & 0.47 & 0.38 & 0.31 & 0.27\\
         \hline
         Spline - 4 Tap & 0.56 & 0.48 & 0.40 & 0.36\\
         \hline
         Spline - 8 Tap & 0.43 & 0.34 & 0.27 & 0.24\\
         \hline
    \end{tabular}}\\
    (b) Delay Accuracy (ns) \\
    \small{
    \begin{tabular}{||c|c|c|c|c||}
        \hline
         Filter & 20\% & 25\% & 30\% & 33\%  \\
         \hline\hline
         Legendre - 4 Tap & 0.338 & 0.254 & 0.198 & 0.172\\
         \hline
         Legendre - 8 Tap & 0.301 & 0.215 & 0.159 & 0.133\\
         \hline
         Spline - 4 Tap & 0.339 & 0.254 & 0.198 & 0.172\\
         \hline
         Spline - 8 Tap & 0.307 & 0.220 & 0.163 & 0.137\\
         \hline
    \end{tabular}}
\end{tabular}
\end{table}

\subsection{Antenna gain  parameterization}

Recall that the antenna responses $G$ are parameterized by a steering angle as well as an interchangeable output or input angle. Since the responses are generally very complicated, the most straightforward approach is to hold samples of $G$ in a look-up table. However, if we want something on the order of angular resolution this table will have $(360\cdot 180)^2$ entries, which would require an exorbitant amount of memory per node. Fortunately, there is a good physical reason for the antenna response to be well structured in a manner we can take advantage of.

Antenna's are generally composed of an array of conducting elements, each of which is has a smooth response as a function of the angle of incidence. For a single element the response is the same for every steering angle\footnote{There are two components to antenna steering: physical and electronic.  Physical steering can be accounted for simply by shifting the arguments into $G$; we will assume that this is already accounted for.  Electronic steering consists of taking a linear combination of the array elements; for a single element, this would be a multiplication by a scalar which can just be incorporated into $G$.} and so we can represent its response as a function $g:\setS^2\mapsto\mathbb{C}$. As these are smooth we can represent them with a relatively small number of $P$ spherical harmonics:
\begin{align}
    \label{eq:sh_antenna}
    g(\theta) = \sum_{p=1}^P b_p \psi_p(\theta) = \vpsi(\theta)^T\vb,
\end{align}
Here $\vpsi(\theta)^T = [\psi_1(\theta),\psi_2(\theta),\dots, \psi_P(\theta)]$ is a vector of evaluations of the spherical harmonic basis functions and $\vb\in\mathbb{C}^P$ are the corresponding coefficients. So for a single element we can represent the response at every angle by a vector of length $P$.
\begin{figure}[t]
    \centering
    \begin{tabular}{cc}
        \includegraphics[width=.4\textwidth]{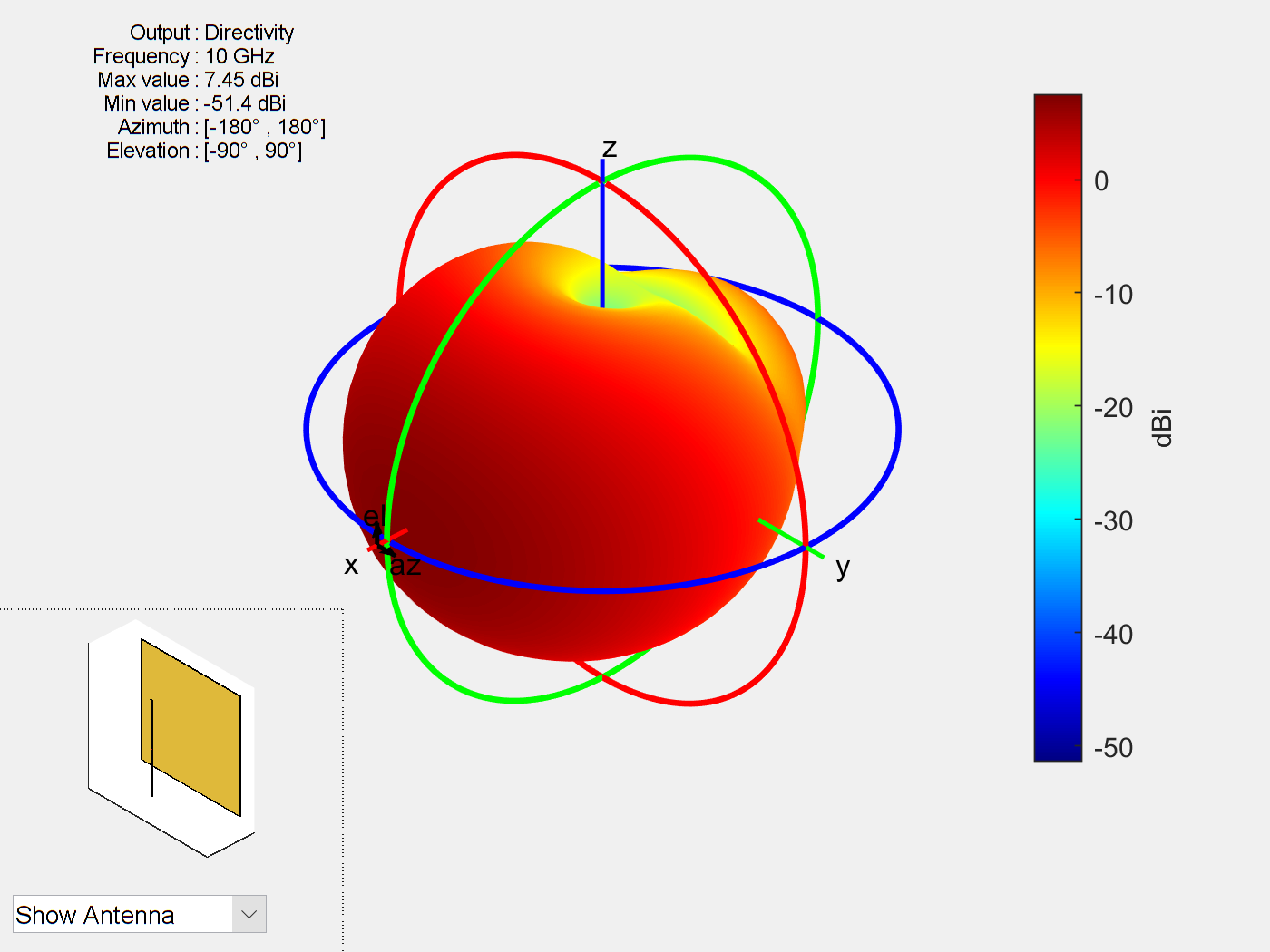} \\ 
        (a) Dipole with reflector\\
        \includegraphics[width=.4\textwidth]{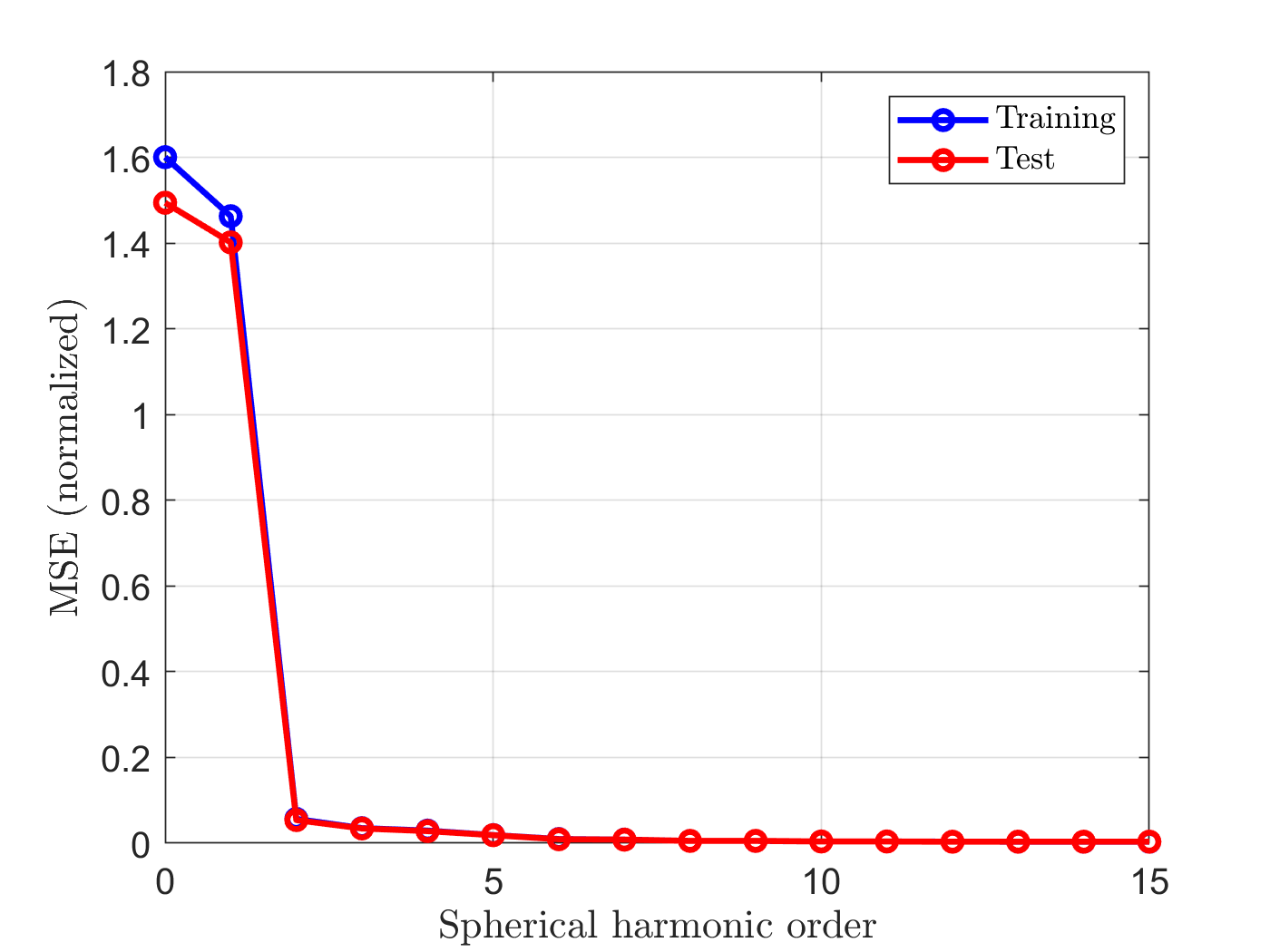}  \\
          (b) MSE vs approximation order
    \end{tabular}
    \caption{\small\sl (a) Radiation pattern of a single dipole with a backplate reflector. The array is composed of $13^2$ antennas of this type arranged in a square pattern.(b) The testing and training MSE of the factorization as a function of the spherical harmonic order. Owing to the smoothness of the dipole response a sharp decays around order 3 (e.g. $P=16$).}
    \label{fig:arrayfit}
\end{figure}

An antenna is composed of an array of $D$ elements and is electronically steered to angle $\theta^s$ by applying a complex-valued set of weights $\vw(\theta^s)\in\C^D$.  Each element $d$ of the array has a different individual response $g_d$.  Additionally each of the elements has a different physical offset from the array center, meaning that this response will be modulated by an angle dependent phase factor $\e^{\j\phi_d(\theta)}$.  The response for the entire antenna (the composite response across all the elements) is 
\begin{align}
    \label{eq:array_factor}
    G(\theta^s,\theta) = \sum_{d=1}^D \bar{w}_d(\theta^s)g_d(\theta)\e^{\j\phi_d(\theta)}
\end{align}	
Thus, the antenna response is an $D$-separable function; any table (matrix) of response values with rows corresponding to $\theta^s$ sampled at discrete points and columns corresponding to $\theta$ sampled at discrete points will have rank at most $D$.

 While we have discussed above the physical reasons that we expect $G$ to be $D$-separable by giving specific roles to its factors, in practice the factorization will be ambiguous.  That is, we simply parameterize the antenna response with $D$ functions $\{g^s_d:\mathcal{S}^2\rightarrow\C\}_{d=1}^D$ and $D$ functions $\{g_d:\mathcal{S}^2\rightarrow\C\}_{d=1}^D$ such that
 \begin{align}
    \label{eq:sh_array}
     G(\theta^s,\theta) = \sum_{d=1}^D \bar{g}_d^s(\theta^s)g_d(\theta).
 \end{align}
 These $2D$ functions themselves are each discretized using spherical harmonics as in \eqref{eq:sh_antenna}.  Thus $G$ is parameterized using $2PD$ numbers $\{\vb^s_d\in\C^P\}_{d=1}^D$, $\{\vb_d\in\C^P\}_{d=1}^D$.

The difficulty of actually fitting a table to \eqref{eq:sh_array} varies greatly depending on our a priori knowledge of the array. Without any knowledge of the array, an explicit rank constraint must be placed on the solution. Though tractable to approximate, it does require a far more complicated approach. Instead, as brief demonstration that the a factorization such as \eqref{eq:sh_array} exists for realistic data, we will fit a model with modest knowledge of the array. We generate a $13\times 13$ uniform planar array composed of dipole antennas with a backplate reflector tuned to $10$ GHz using the MATLAB antenna toolbox as shown in Figure~\ref{fig:arrayfit}(a). We make the mild assumption that we know the geometry of the array and that the steering weightings are of the form $\vw(\theta^s)^T = [w_1\e^{j\phi_1(\theta^s)},w_2\e^{j\phi_2(\theta^s)},\dots, w_D\e^{j\phi_D(\theta^s)}]$. We sample $\theta,\theta^s$ at $3500$ points respectively to generate a $3500 \times 3500$ response table. This table was then split $9:1$ into training and test data sets. Following a least squares fit the normalized MSE as a function of the spherical harmonic order is presented in Figure~\ref{fig:arrayfit}(b). The similar trends in the test and training data suggests the model generalizes well, meaning we can safely evaluate arbitrary angles with this model. The parameterization, even for this relatively small antenna table, yields an immense amount of data compression. The entire $3500 \times 3500$ table can be represented by a 5408 parameters (for an order 3 approximation). This reduces the amount of data we have to store to $\approx 0.05\%$ of the nominal table size.

\subsection{Scattering profiles}
\begin{figure}[h]
    \centering
    \begin{tabular}{cc}
        \includegraphics[width=.4\textwidth]{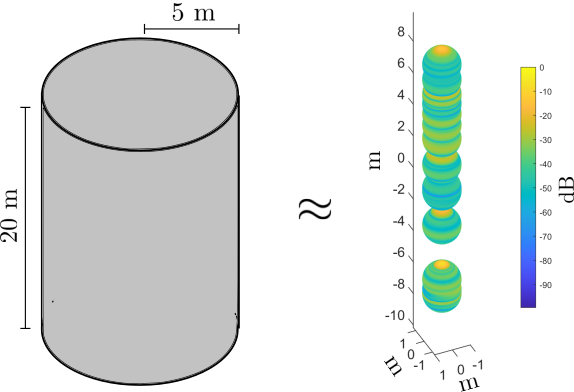} \\
        (a) Cylinder and scattering approximation\\
        \includegraphics[width=.4\textwidth]{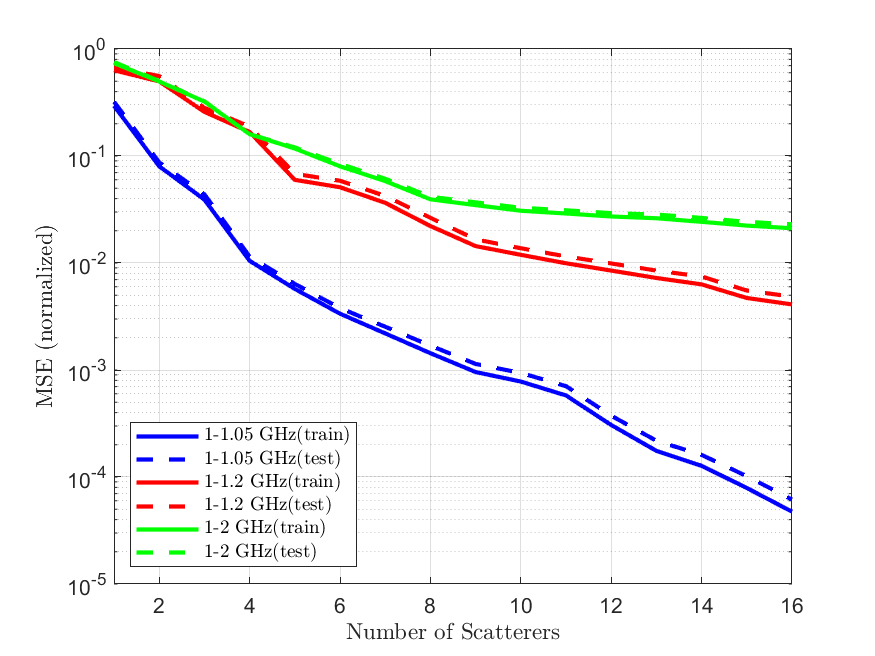}  \\
          (b) Scattering model MSE
    \end{tabular}
    \caption{\small\sl (a) An electrically large (e.g.\ $\gg\lambda$) metallic cylinder's monostatic scattering response is fit to a $K$-point anisotropic scattering model. (b) We fit each scattering point with order 15 (e.g.\ $P=256$) spherical harmonic basis functions, and observer the mean squared error (MSE) for a varying number of scatterers and observation bands.}
    \label{fig:cylinder}
\end{figure}
As discussed in Section~\ref{sec:directpath} the ability to form a separable scattering model as in \eqref{eq:pointscatteringsep} is critical to the computational and memory improvements of the direct path model. This is straightforward to produce for isotropic scattering models, but a richer model that can represent more complicated responses involves fitting a bilinear response model. 

This paradigm is not too dissimilar from the above discussion on antenna representations. In particular we assume that the input and output scattering responses in \eqref{eq:pointscatteringsep} can be represented as 
\begin{align*}
    \alpha_k(\theta^i) = \vpsi(\theta^i)^T\vb_k^i,~\beta_k(\theta^o) = \vpsi(\theta^o)^T\vb_k^o,
\end{align*}
where $\vpsi(\theta)$ is defined as in \eqref{eq:sh_antenna}. So the relative delays $\{\tau_k^i\}_{k=1}^K,\{\tau_k^o\}_{k=1}^K$ account for the relative offset of the scatters while the above parameterization accounts for the bistatic anisotropic response of each point. So with a set of locations $\{\vx_k\in\R^3\}_{k=1}^K$ and a sets of coefficients $\{\vb^i_k\in\C^P\}_{k=1}^K$, $\{\vb_k^o\in\C^P\}_{k=1}^K$ we can approximate the scattering profile at any angle, requiring us to store $K(2P+3)$ numbers.

Fitting this model to bistatic scattering measurements from an object is difficult, but tractable. \cite{huang2021an,huang2022sp} suggests using the monostatic response, e.g.\ where $\theta^i=\theta^o$, to solve for the point locations over a fixed grid using a modified orthogonal matching pursuit (OMP) algorithm. With these locations in hand they then propose finding the bistatic response by formulating it as a bilinear inverse problem that is solved using a normalized iterative algorithm. Their results show that this method can find a reasonable fit to the bistatic scattering response of complicated objects, such as a commercial aircraft. 

As a demonstration of this concept we fit an anisotropic scattering model to an electrically large (e.g.\ it is far greater in size than the probing signal's wavelength) metallic cylinder. Data on the scattering profile was generated using CST, which probed the object at varying angles over a frequency band of 1 to 2 GHz in 50 MHz steps\footnote{This was empirically found to be sufficient spacing for capturing the frequency variation.}. The data, collected at angular resolution, amounted to a $21\times 180 \times 360$ tensor or $1360800$ data points. This was then split in a $9:1$ ratio into training and test data. The search space for scatterer locations was set to be grid points in $10 \times 10 \times 20 $ m block, spaced uniformly by $\lambda/2$ where the wavelength was set according to the maximum frequency (e.g.\ $2$ GHz). To fit the monostatic model we use a modified version of the OMP style algorithms in \cite{huang2021an,huang2022sp}. The modification being that augmented the algorithm with the fast non-uniform Fourier and spherical harmonic techniques given in \cite{keiner2009us} to make it computationally efficient, which was a necessity for this scale of data. Figure~\ref{fig:cylinder}(a) provides a qualitative example of what these models look like, and unsurprisingly the ``learned" scattering locations are in a line along the major axis of the cylinder.

We fit the model to a varying number of scattering points and observation bands, and observe the mean squared error (MSE) for test an training points. The spherical harmonic depth was fixed to be order 15 such that $P=256$. The normalized MSE for a few bands vs the number of scattering points used in the fit are given in Figure~\ref{fig:cylinder}(b). The trends of these results indicate that, as one would assume, increasing the number of scattering points monotonically decreases the error. Predictably, the quality of the approximation is dependent on the bandwidth of the probing frequencies. At low bandwidths such as $50$ MHz the model is exceptionally good, achieving nearly $3$ digits of precision. For the full $1$ GHz bandwidth the error is larger but has an absolute error within $\approx 10\%$ of the true response. If this is considered within an acceptable tolerance, then this model can represent the full $1360800$ data points by a modest $4096$ parameters, equating to an immense amount of memory savings. If we wish to drive this model to a smaller error, then we can simply increase the spherical harmonic depth or number of scattering points.
\begin{figure}[h]
    \centering
    \begin{tabular}{cc}
        \includegraphics[width=.4\textwidth]{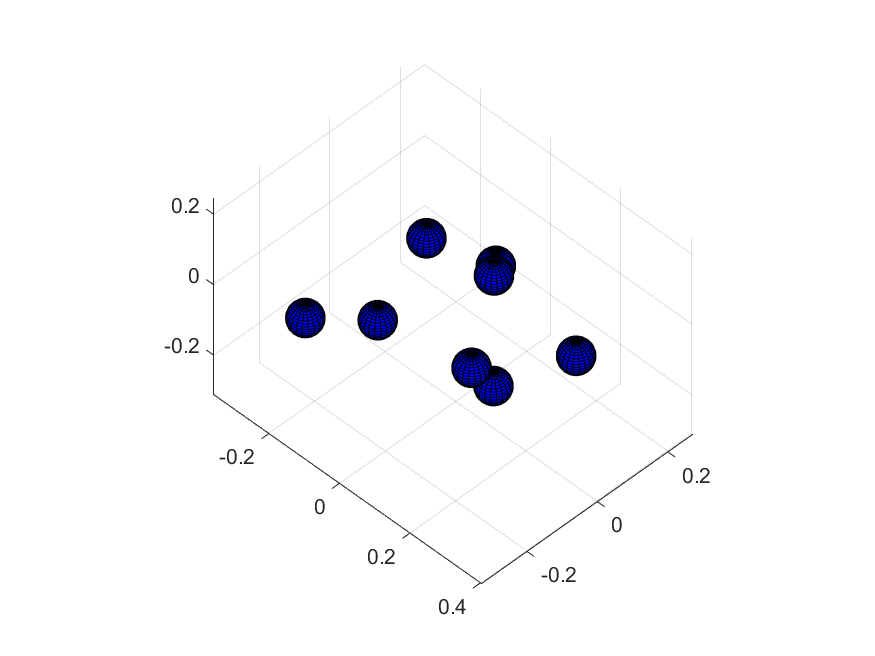}\\
        (a) Isotropic Scattering Points\\ 
        \includegraphics[width=.4\textwidth]{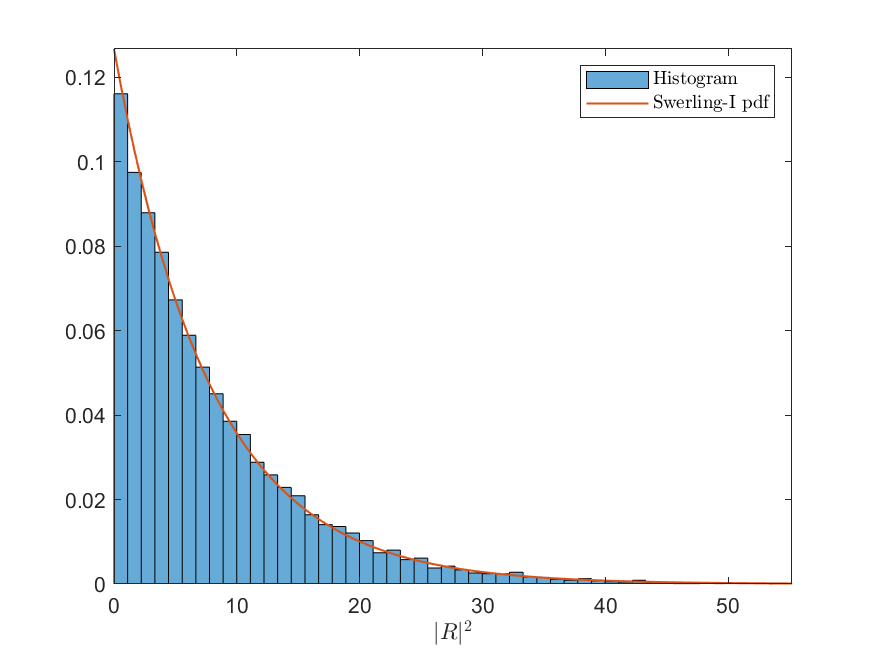}  \\
         (b) Matched filter response histogram
    \end{tabular}
    \caption{\small\sl (a) Isotropic scatterers are placed in a large ($\gg \lambda$) region, approximating the response of a large object. (b) The scattering model is randomly rotated and illuminated by a signal. The magnitude of the matched filter response fluctuates randomly, producing the displayed histogram. The red line indicates the expected distribution of this model.}
    \label{fig:swerling1}
\end{figure}

It is worth noting that in addition to being suitable for modeling complex object geometries the point scattering model envelops random fluctuation loss models in a natural way. Fluctuation models, often labeled ``Swerling" models, model the response of a radar to a moving target as a random variable with some suitable distribution for the scenario \cite{swerling1960pr}. These models themselves are derived from the assumption that an objects scattering response behaves like a set of isotropic points. Essentially the type of distribution is dependent on the distribution of the isotropic weights e.g.\ Swerling 1 assumes equal weiightings while Swerling 3 assumes one weight is dominant. The ``randomness" comes from the summation of these responses fluctuating as the object moves and rotates due to constructive and deconstructive interference. Hence, we can model a fluctuating response by simply using a suitable isotropic scattering model and rotating the object throughout emulation. 

As a brief demonstration of this concept we placed several isotropic scattered at random locations in an large (relative to $\lambda$) area as shown in Figure~\ref{fig:swerling1}(a). We then illuminated the model with a windowed sinusoidal signal as the object randomly rotated, meeting the requirements for a Swerling 1 hypothesis. A histogram of the matched filter response's fluctuations is shown in Figure~\ref{fig:swerling1}(b) and very closely matches the expected Swerling 1 distribution.

\section{Numerical experiments}
\label{sec:experiments}

The discussions above established the direct path model for RF emulation, as well as how to leverage it to greatly increase computational and memory efficiency. However, their still remains a question about the ability to practically implement the model and it's effectiveness under realistic computational constraints. For instance the accumulation of rounding errors in various fixed and floating point operations can greatly impact emulation quality. These effects have been well studied in TDL model, but are not as well understood in the direct path model. In this section we pacify these concerns by testing our model in a variety of scenarios using a cycle-level simulator of an actual hardware implementation. In particular, we focus on scenarios where we can analytically calculate a baseline for comparison to simulator outputs. This way we can observe the accuracy of the model while it is subject to realistic computational non-idealities.

Each test uses a modified version of the cycle-level C++ simulator used extensively in \cite{mukherjee2023ah} and part II to test the ASIC implementation. The only major modification is that this simulator is meant to test the full-scale version of the emulator, where objects have more than one scattering point and we are not limited to only $4$ objects in a scenario. Objects can have up to $K=16$ anisotropic scattering points, with each point allows to have $P=256$ spherical harmonic expansion coefficients for the incoming and outgoing scattering response. It operates at a sampling frequency of $f_s=2.5$ GHz with a maximum signal frequency of $2$ GHz, meaning it is $25\%$ oversampled. Owing to this degree of oversampling, it uses $R=4$ tap fractional delay filters. The maximum operating range is $500$ km, meaning samples from $v_{m,k}(t)$ are places in a buffer of length $L\sim 8.3 \cdot 10^6$. Finally the simulator updates the scenario parameters every $1.3$ ms, meaning that position, velocity, orientation, etc. are updated at this rate. The simulator assumes baseband processing, meaning the signal has been mixed down from its center frequency $f_c$. The $f_c$ is kept as an additional parameter such that Doppler can be properly compensated. Finally, we note that the simulator assumes processing has been distributed, meaning each node operates as an independent piece of hardware and send signal to other nodes via a series of communication network links.

\subsection{Interferometry}
\begin{figure}[t]
    \centering
    \begin{tabular}{cc}
        \includegraphics[width=.4\textwidth]{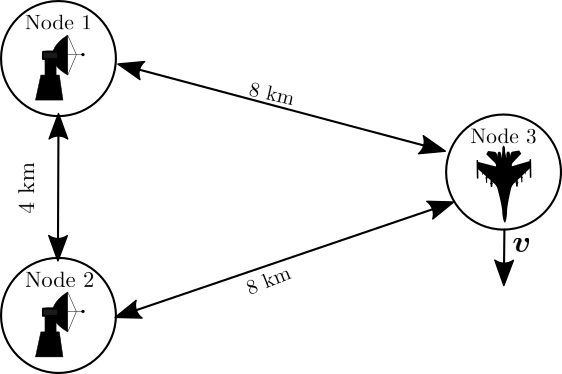}\\
        (a) Scenario diagram\\
        \includegraphics[width=.4\textwidth]{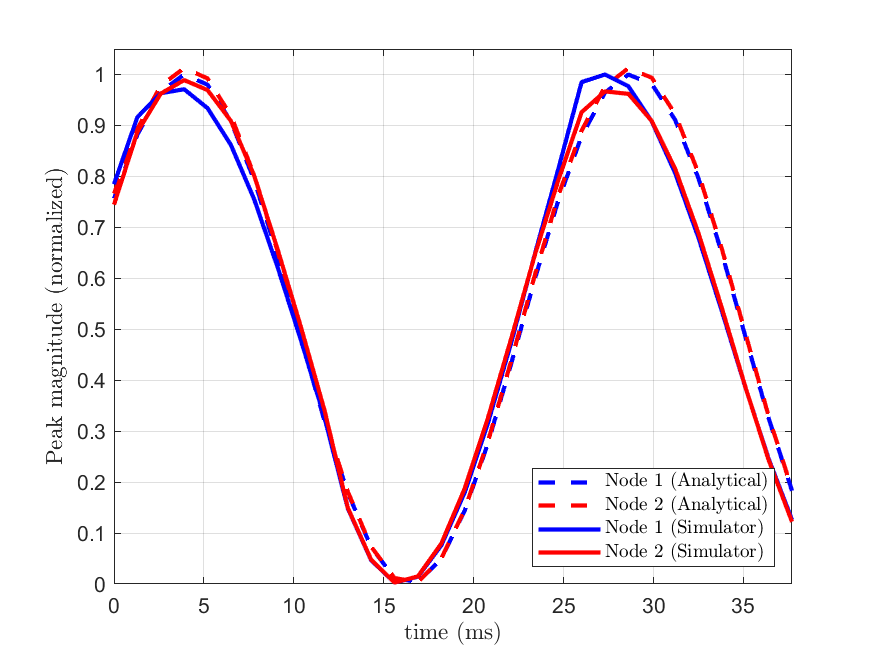}  \\
         (b) Matched filter peaks
    \end{tabular}
    \caption{\small\sl (a) Symmetry scenario 1 contains two transmitters/receivers, spaced 4 km apart, illuminating a slow moving target. The target moves parallel to the axis of separation between the transmitter/receiver pair. (b) The normalized magnitude of the matched filter response produced by node 1 and node 2 over time. The dotted lines represent the analytically calculated response while the solid lines are the simulator outputs. }
    \label{fig:interferometry}
\end{figure}
Our first scenario is tests the dynamic modeling capabilities of the direct path model. The scenario, depicted in Figure~\ref{fig:interferometry}(a), places two nodes acting as transmitters and receivers 4 km apart. A third node, acting as pure reflector is put in an initial position 8 km from the both of the other nodes. The reflector then moves parallel to the axis of separation at a speed of $100$ m/s. Based on the relative spacing between nodes the superposition of reflections will either constructively or destructively interfere. Furthermore, since the object is moving the degree of interference will vary with time, leading to a predictable ``beat" pattern. Hence we can verify the dynamic modeling capability by collecting responses from the receive nodes and observing this pattern.

The transmitters both isotropically radiate pulses composed of a $1$ GHz sinusoid windowed to $20$ $\mu$s. To prevent self-interference the antenna gains from node 1 to node 2 are set to 0, and similarly we set the scattering responses of these nodes to 0 as well. The scattering profile of node 3 is set to a single isotropic scattering point. Each pulse return is matched filtered, and we collect these responses over time to observe the beat pattern. As a baseline, we analytically calculate the expected response as well. As shown in Figure~\ref{fig:interferometry}(b), the simulators response closely matches the analytical response.

\begin{figure}[h]
    \centering
    \begin{tabular}{cc}
        \includegraphics[width=.4\textwidth]{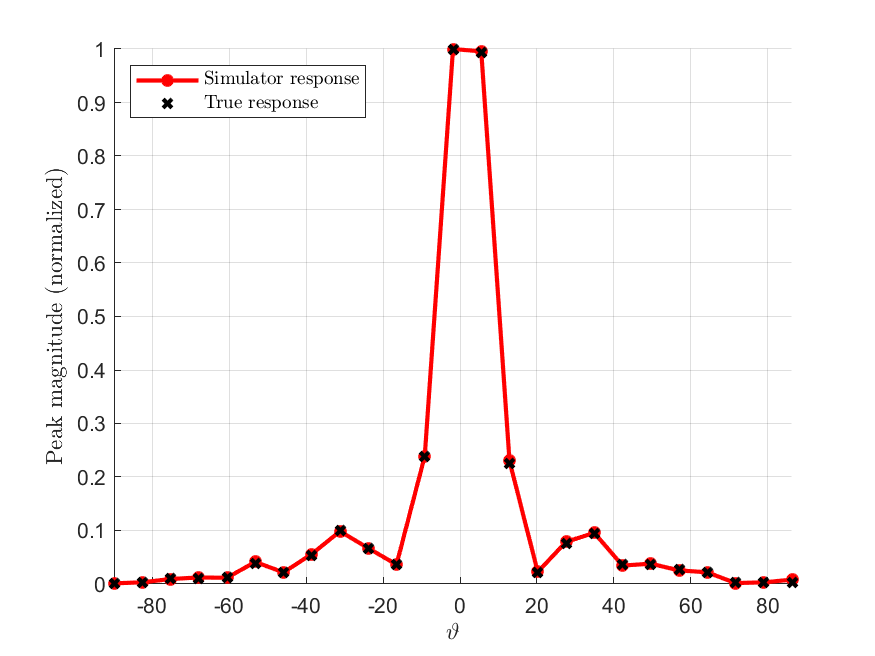}\\
        (a) $\theta^s = (3.67^o,1.83^o)$\\
        \includegraphics[width=.4\textwidth]{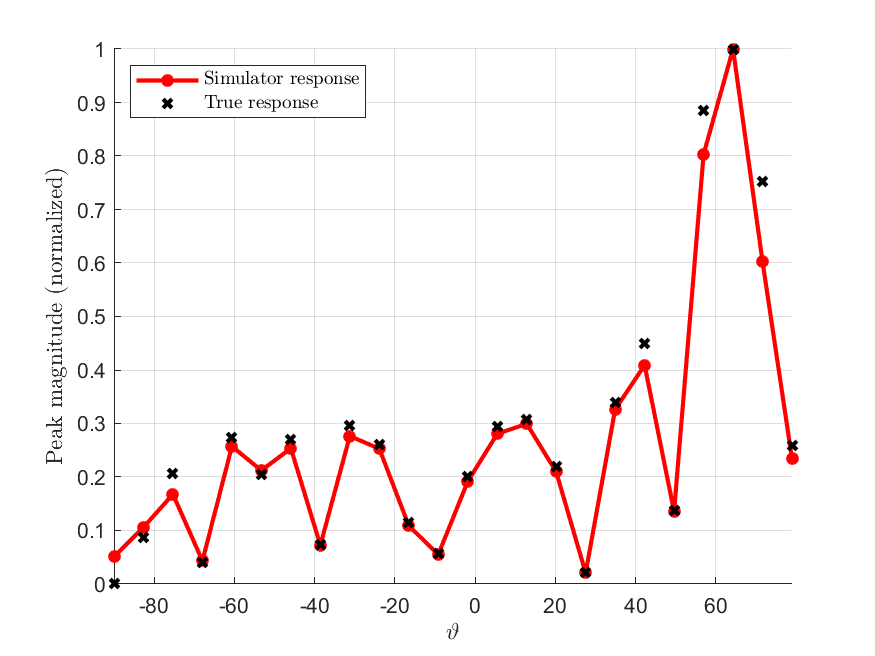}  \\
          (b) $\theta^s = (3.67^o,69.7^o)$
    \end{tabular}
    \caption{\small\sl A receive node traverses a semicircular path around a transmitter node from $\vartheta=90^o$ to $\vartheta=-90^o$ for $\varphi = 0^o$. (a) For one pass the transmitter node's array is steered to $\theta^s = (3.67^o, 1.83^o)$ and in another (b) it is steered to $\theta^s = (3.67^o, 69.7^o)$. The receive node's matched filter response produces by the simulator is given by the red dots and the analytically calculated response is given by the black x's. }
    \label{fig:antennaresp}
\end{figure}

\subsection{Antenna responses}
In our second test scenario we verify that the model can accurately represent a complicated antenna response in it's factored form. For this we require two nodes; a transmitter and a receiver. The transmitter is given the $13 \times 13$ UPA array response we previously factored in Section~\ref{sec:techcon}. The array is then electronically steered as it emits a windowed $10$ GHz sinusoid. The receiver node then traverses a semi-circular path from $\vartheta=90^o$ to $\vartheta=-90^o$ for $\varphi = 0^o$ relative to the transmit node. By measuring the magnitude of the matched filter response at the receive node as a function of angle we can essentially test that the correct beam pattern is observed.

\begin{figure}[t]
    \centering
    \begin{tabular}{cc}
        \includegraphics[width=.4\textwidth]{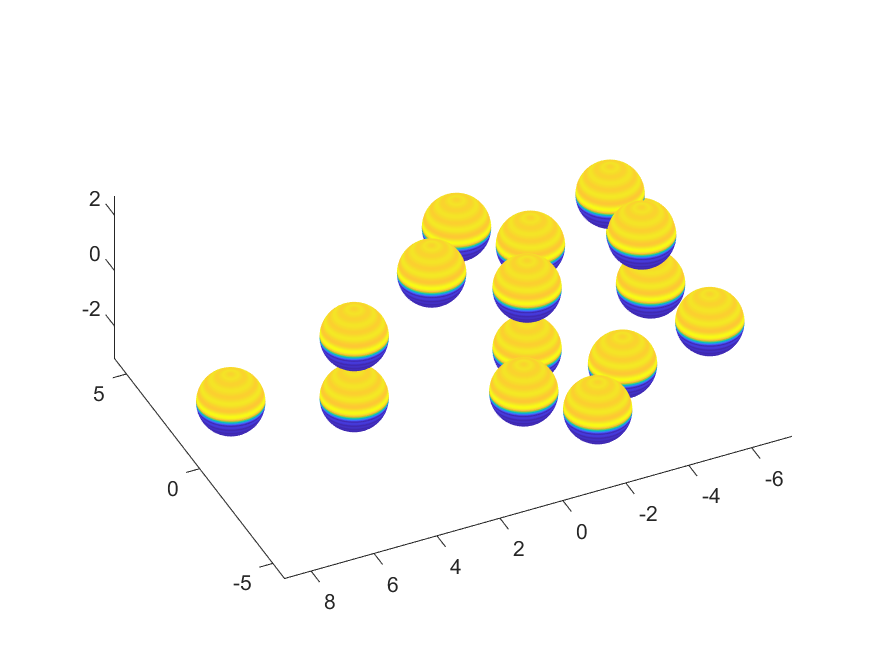}\\
        (a) Anisotropic scattering profile\\
        \includegraphics[width=.4\textwidth]{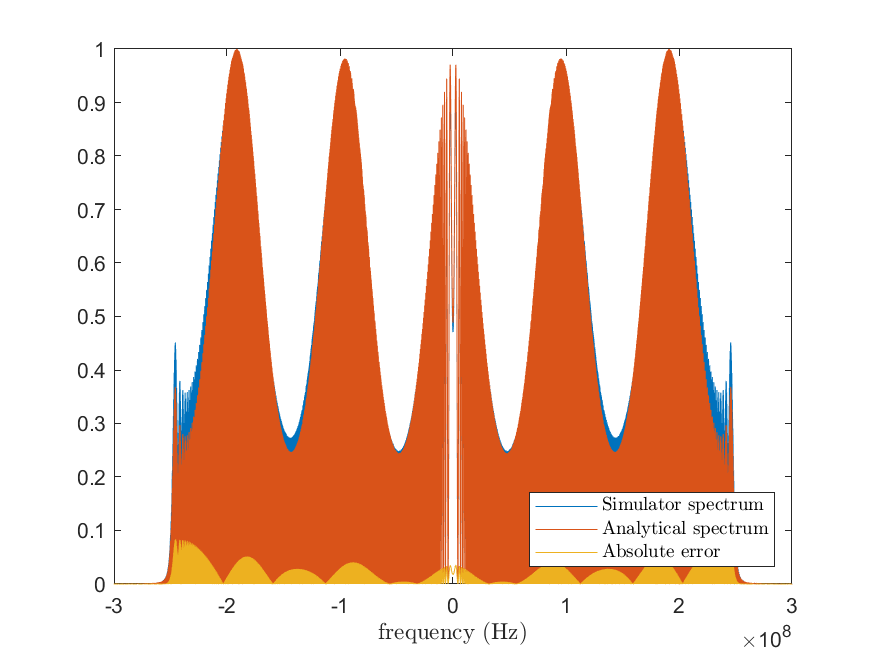}  \\
          (b) Normalized response periodogram 
    \end{tabular}
    \caption{\small\sl (a) Several anisotropic scatterers, each modeling a spherical harmonic approximation to a spherical heaviside response, are placed in a region geometrically resembling a plane. (b) The model is illuminated with a 500 MHz chirp, and the spectrum of the received signal is generated. We compare this to the analytically calculated response to the signal, and observe a small relative error between the two.}
    \label{fig:complexscattering}
\end{figure}

The receiver traverses this path twice, with an electronic steering angle of $\theta^s = (3.67^o,1.83^o)$ for the frist pass and $\theta^s = (3.67^o,69.7^o)$ for the second. As a baseline we analytically calculate the matched filter response of the signal at each angle using the gains given by the MATLAB antenna simulator. The simulator responses along with the analytically calculated responses at these steering angles are shown in Figure~\ref{fig:antennaresp}. It is clear from these results that the simulator and its factored antenna response well represents the expected response.

\subsection{Complex scattering profile}
The last test scenario is designed to examine the simulators ability to model a complicated scattering profile. To do this we require two nodes; one acting as a transmitter/receiver and the other acting as a pure reflector. To generate the reflector we place a series of anisotropic scattering points in a pattern reminiscent of plane as shown in Figure~\ref{fig:complexscattering}(a). The anisotropic pattern of each is set to be the bistatic spherical harmonic approximation to a heaviside function. This pattern was chosen because it is more easily verified and interpretable than an arbitrary pattern.

The transmitter node illuminated the reflector node with a $500$ MHz LFM at a relative angle of $\theta = (0^o,0^o)$. The signal received at the receiver node was then discrete Fourier transformed (DFT). As a baseline the analytical response of the signal to this scattering profile was calculated and used to generate a complimentary DFT spectrum. Periodograms of the simulated and analytically calculated response, and the absolute error between the spectrums, are shown in Figure~\ref{fig:complexscattering}(b). These results show a small relative absolute error between the simulated output its anyltical counterpart. The source of this error is likely due to minor inaccuracies in the fractional delay filters coupled with an accumulation of rounding errors. We note that the response is clearly behaving like a FIR filter as predicted since the response of an LFM is spectrally flat with only minor ripples.

\section{Conclusion}
\label{sec:conclusion}

In part I of this series we developed a new ``direct path" computational model for real-time digital RF emulation. We showed, through a careful mathematical formulation and simulation, that this model can suitably emulate all channel characteristics necessary for RF system testing. Furthermore, by leveraging mild assumptions on the physical characteristics of scattering profiles and antenna structures our model was shown to yield tremendous computational benefits. This, coupled with a naturally distributed framework, motivated us to explore hardware implementations of the model.

Part II of this series focused on the development of working implementations of the direct path computational model. We approached this from two design perspectives, the first being an ASIC that leverages near-memory computations and autonomous distributed control. This allowed us to achieve high bandwidth and low-latency performance that is not viable in off-the-shelf component based systems. The second design used an FPGA to to implement a larger scale (e.g.\ greater number of objects) system. Though this style of implementation operates at a lower bandwidth than the ASIC version, it demonstrates that the model can be viably scaled to incorporate more objects as needed.

The complementary results of these two papers have, together, established a new and interesting option for consideration in future development of RF emulators. Through our collaborative effort we have coupled innovative modeling techniques with cutting-edge high performance computing paradigms. The result being a series of implementations that operate at a level of performance not achievable by more traditional designs, and a model that can be efficiently scaled to meet testing requirements.   

\bibliographystyle{IEEEtran}
\bibliography{IEEEabrv.bib,directpathrefs.bib}

\appendix
\section{Approximations to relative motion}
\label{sec:apdx_doppler}

For this discussion we require some model of the scenarios dynamics. Though this can be done in several ways a simple model is to assume that over a short timescale the motion is linear. To formalize this let $\vx_n(t)$ be the time varying position of the $n$th object. Fix a time $t_0$ and let $\vx_n(t_0)$ and $\vv_n(t_0)$ be the position and velocity at this time. Then for $t \in [-\epsilon,\epsilon]$, with $\epsilon$ suitably small, the position of the object is modeled as
\begin{align*}
    \vx_n(t) = \vx_{n}(t_0) + t \vv_{n}(t_0).
\end{align*}
With this in hand we model the distance between two arbitrary nodes $n$ and $n'$ as
{\small
\begin{align*}
    \| \vx_{n'}(t)-\vx_{n}(t) \|_2 &= \| \vx_{n'}(t_0)-\vx_{n}(t_0) +t(\vv_{n'}(t_0)-\vv_{n}(t_0)) \|_2\\
    &= \| \vd_{n,n'} + t\vv_{n,n'}\|_2,
\end{align*}}
and expanding on this
{\small
\begin{align*}
    \| \vd_{n,n'} + t\vv_{n,n'}\|_2&= \\
    &\left(\|\vd_{n,n'}\|_2^2 + 2t\< \vd_{n,n'},\vv_{n,n'}\> + t^2 \|\vv_{n,n'}\|_2^2 \right)^\frac{1}{2}.
\end{align*}}
If we assume $\|\vd_{n,n'}\|_2\gg t \|\vv_{n,n'}\|_2$, which in general is true, then
{\small
\begin{align*}
   &\left(\|\vd_{n,n'}\|_2^2 + 2t\< \vd_{n,n'},\vv_{n,n'}\> + t^2 \|\vv_{n,n'}\|_2^2 \right)^\frac{1}{2} \approx \\
   &\qquad\qquad\qquad\qquad\qquad\left(\|\vd_{n,n'}\|_2^2 + 2t\< \vd_{n,n'},\vv_{n,n'}\>  \right)^\frac{1}{2}.
\end{align*}}
Going further
{\small
\begin{align*}
   &\left(\|\vd_{n,n'}\|_2^2 + 2t\< \vd_{n,n'},\vv_{n,n'}\>  \right)^\frac{1}{2} = \\
   &\qquad\qquad\qquad\qquad\|\vd_{n,n'}\|_2 \left(1 + 2t\frac{\< \vd_{n,n'},\vv_{n,n'}\>}{\|\vd_{n,n'}\|_2^2}  \right)^\frac{1}{2}\\
   &\qquad\qquad\qquad\qquad\approx \|\vd_{n,n'}\|_2 \left(1 + t\frac{\< \vd_{n,n'},\vv_{n,n'}\>}{\|\vd_{n,n'}\|_2^2}  \right)\\
   &\qquad\qquad\qquad\qquad= \|\vd_{n,n'}\|_2  + t\frac{\< \vd_{n,n'},\vv_{n,n'}\>}{\|\vd_{n,n'}\|_2},  
\end{align*}}
where the second line follows from a Taylor expansion. We note that $\frac{\<\vd_{n,n'},\vv_{n,n'}\>}{\|\vd_{n,n'}\|_2} = \|\vv_{n,n'}\|_2\cos\theta$ where $\theta$ is the angle between $\vd_{n,n'}$ and $\vv_{n,n'}$. Hence the relative variation in delay is only a function of the relative \emph{radial} velocity of the objects. With this approximation in hand the delay between nodes $n$ and $n'$ can be applied to the receive signal as 
\begin{align*}
    &s \left (t-\frac{1}{c}\|\vd_{n,n'}\|_2  - \frac{t}{c}\frac{\<\vd_{n,n'},\vv_{n,n'}\>}{\|\vd_{n,n'}\|_2}  \right ) =\\
    &\qquad\qquad\qquad\qquad s \left (t-\tau_{n,n'}  - \frac{t}{c}\frac{\<\vd_{n,n'},\vv_{n,n'}\>}{\|\vd_{n,n'}\|_2}  \right ).
\end{align*}
The time dependence of the second delay term means that the signal is being dilated in time, and hence is no longer LTI. For simplicity let $\rho_{n,n'} = \frac{t}{c}\frac{\<\vd_{n,n'},\vv_{n,n'}\>}{\|\vd_{n,n'}\|_2}$ such that the above can be expressed as $s(t-t\rho_{n,n'}-\tau_{n,n'})$. Let $s(t)$ be an $\Omega$-bandlimited signal $u(t)$ modulated to a center frequency $f_c$ such that $s(t) = e^{j2\pi f_c t}u(t)$ is supported over the frequency interval $[f_c-\Omega,f_c+\Omega]$. Without loss of generality, and for brevity, let $tau_{n,n'}=0$ so that we can concentrate on the $\rho_{n,n'}$ term. Then,
\begin{align*}
    s(t-t\rho_{n,n'}) = e^{j2\pi f_c (1-\rho_{n,n'}) t}u(t(1-\rho_{n,n'}) )
\end{align*}
and if we assume the signal has been demodulated by $e^{-j2\pi f_c t}$ for baseband processing, then
{\small
\begin{align*}
    \hat s(t-t\rho_{n,n'}) &= e^{-j2\pi f_c t}s(t-t\rho_{n,n'})\\
    &= e^{-j2\pi f_c \rho_{n,n'} t}u(t(1-\rho_{n,n'}) ).
\end{align*}}
The effects of the dilation or compression term $1-\rho_{n,n'}$ on $u(t)$ is generally considered negligible over the band $[-\Omega,\Omega]$ so 
\begin{align*}
    \hat s(t-t\rho_{n,n'}) \approx e^{-j2\pi f_c \rho_{n,n'} t}u(t ).
\end{align*}
This approximation will degrade as the bandwidth and/or velocity increase, but for most ranges of interest it is perfectly sufficient.

\end{document}